# The possible superconductivity at 109 K in YBaCuO materials

P.Udomsamuthirun[1,2], T.Kruaehong[1], T. Nilkamjon[1,2] and S. Ratreng[1,2]

[1] Prasarnmitr Physics Research Unit, Department of Physics, Faculty of Science, Srinakharinwirot University, Sukumvit 23, Bangkok, 10110, Thailand.
E-mail: udomsamut55@yahoo.com

[2] Thailand Center of Excellence in Physics(ThEP), Si Ayutthaya Road, Bangkok, 10400, Thailand.

**Abstract**

The new YBaCuO superconductors are synthesized by using the standard solid state reaction method as Y5-6-11, Y7-9-16, Y5-8-13, Y7-11-18, Y156, Y3-8-11, and Y13-20-23. We find that all material obtained are shown the Meissner effect at 77 K. The resistivity measurements are used by the four-probe method. The Y 7-11-18 has the highest $T_c$ onset as 109 K. The XRD spectra are shown that they have the same crystal structure as Y123 with some impurities peaks.

**1. Introduction**

In 1986 Bednorz and Muller [1] found the first high temperature superconductors in La214 compound that showed the transition temperature ($T_c$) above 30 K. And in 1987 the transition temperature of $YBa_2Cu_3O_7$ (Y123) was increased to around 92 K by Chu and coworkers[2]. The researchers have been carried out on the YBaCuO-family compound like Y123, $YBa_2Cu_4O_8$ (Y124), and $Y_2Ba_4Cu_7O_{15}$ (Y247) to find a higher $T_c$ in the YBaCuO-family. They found that Y124 and Y247 became superconductor at 80 K[3] and 40K[4], respectively. The Y247 exhibits a superconducting transition with $T_c$ ranging from 30 to 95 K, depending on the oxygen content[5,6].

Recently Aliabadi, Farshchi and Akhavan [7] found the new Y-based high temperature superconductor in $Y_3Ba_5Cu_8O_{18}$ (Y358) that become superconductor at 102 K. And they also proposed that in order to have a stronger superconductor with higher $T_c$ in YBaCuO-family one should pump more holes from the chains to the oxygen sites of the planes-tending to diagonal charge order. The Y123 has two $CuO_2$ planes and one $CuO$ chain. The Y124 has one $CuO$ double chain. Y247 has one $CuO_2$ planes and one $CuO$ chain, and one double chain. Y358 has crystal structure similar to Y123 with five $CuO_2$ planes and three $CuO$ chains. The increasing in the number of $CuO_2$ planes and $CuO$ chain have important effect on the $T_c$ of Y358. The YBaCuO-family have shown the different in their number of $CuO_2$ planes and $CuO$ chains or double chains that believed to be the carrier reservoirs. However, Nakajima et al.[8] had proposed the limited increase in the number of the $CuO_2$ planes in all high-$T_c$ cuprate superconductors to three.

We think that there should be the relations between the superconductors in YBaCuO-family. The assumptions about the relation in these material are made and we synthesize the new superconductors in this family by using our assumptions . We can find the new YBaCuO superconductors in 7 formula with the difference in critical temperature.

**2. Experimental assumptions**

We know that the YBaCuO-family are consists of Y123,Y124,Y247 and Y358 with the Y358 is the highest $T_c$ of this family . The Y123 and Y358 are shown the similar crystal structure[7] . Aliabadi,Farshchi and Akhavan [7] proposed that the lattice parameters, $a$ and $b$ , of Y123 is very close to of Y385 but the lattice parameters, $c$ ,of Y358 is almost 3 time of Y123 . Y123 has two $CuO_2$ planes and one $CuO$ chain and Y358 have five $CuO_2$ planes and three $CuO$ chains. Y358 have five $CuO_2$ planes that 2.5 time of the $CuO_2$ planes of Y123 . We think that three parameters should have some relations as

1.the number of $CuO_2$ planes and number of Ba-atom

2.the number of $CuO$ chains and the number of Y-atom.

3.the number of Ba-atom plus Y-atom are equal to the number of Cu-atom.

The relation between the number of $CuO_2$ planes and number of Ba-atom, and the number of $CuO$ chains and the number of Y-atom can not prove in this paper. However, the number of Ba-atom plus Y-atom equal to the number of Cu-atom can be done by the experimental. In Y123, there is 1 Y-atom and 2 Ba-atoms so we get 3 Cu-atoms. In Y358, there are 3 Y-atom and 5 Ba-atoms so we get 8 Cu-atoms. So we think that the main ideal to synthesize a new superconductor in this family is the number of Ba-atom plus Y-atom equal to the number of Cu-atom.

To reach the highest $T_c$ , we should pump more holes in this family. As our relation that the number of $CuO_2$ planes relate to number of Ba-atom . We need more $CuO_2$ planes so we will not do anything to Ba-atom . To make holes, the number of Y-atom should be missing. This concept are agreed with the assumption to synthesize Y123 that replacing the La-atom by Y-atom in $BaCuO_2$ perovskite; $Y^{3+}$ has a radius smaller than $La^{3+}$ ; and the $T_c$ is higher. We make the assumption that the number of Y-atom should be missing to create more holes to higher the critical temperature. .

At this point, we can make the assumptions to synthesize a new superconductors in the YBaCuO-family as

1.the number of Ba-atom plus Y-atom are equal to the number of Cu-atom.

2.the number of Y-atom can be missing to reach the higher $T_c$ but the 1$^{st}$ assumption must be obeyed.

The Y123 and Y358 can be explained by our assumption as Y123 is no Y-atom missing and Y358 is 1 Y-atom missing every 5 Ba-atom.

According to our assumptions, there are many new superconductors will be found. Example, In case of 1 Y-atom missing ,the general formula should be $Y_{x-2}Ba_xCu_{2x-2}O_\delta$ . The Y358 is the example of this group which has the percent of Y-atom missing to number of Ba-atom as $\frac{1}{5}x100 = 20\%$ . The 2 Y-atom missing, the general formula should be $Y_{x-3}Ba_xCu_{2x-3}O_\delta$. We can get Y5-8-13 with the percent of Y-atom missing to number of Ba-atom as $\frac{2}{8}x100 = 25\%$.

### 3. Experimental details

To prove our assumptions, we synthesize a new group of YBaCuO superconductors by using the standard solid state reaction method. Appropriate stoichiometric ratios of powder of $Y_2O_3$, $BaCO_3$ and $CuO$ are mixed, ground, and react in air at $950^0C$ for 24 hour, and cool to 100 $^0C$. Calcinations is repeated twice with intermediate grinding. The powders are reground, pressed into pellets 30 mm in diameter and about 5 mm thickness under 2000 psi pressure. Finally, the samples obtained are sintering at $950^0C$ for 24 hour and annealing at 500 $^0C$ for 24 hour in air. At this point we do not interest in the effect of oxygen-doping so we obtain all sample in air annealing.

We firstly test the superconductivity state by using the Meissner effect at 77 K and find that all material obtained are shown the Meissner effect as Table 1. These mean that all of our samples are superconductors with the critical temperature above 77 K.

| Compound | Y-atom missing per Ba-atom | % of missing | Shown Meissner effect at 77 K |
|---|---|---|---|
| Y123 | 0 | 0 | Yes |
| Y5-6-11 | 0 | 0 | Yes |
| Y7-9-16 | 1:9 | 12.5% | Yes |
| Y358 | 1:5 | 20% | Yes |
| Y5-8-13 | 2:8 | 25% | Yes |
| Y7-11-18 | 3:11 | 27% | Yes |
| Y156 | 3:5 | 60% | Yes |
| Y3-8-11 | 4:8 | 50% | Yes |
| Y13-20-33 | 6:20 | 30% | Yes |

Table 1. Shown the new YBaCuO -superconductors synthesized.

The resistivity measurements are used by the four-probe method. All samples show the fact that with increasing measuring current the onset of resistivity drop are shifted to lower temperature. The current densities $J = 2.55x10^3$ A/m$^2$ used are shown in Figure 1. And the normalized resistivity versus temperature are shown in Figure 2. The summation of the $T_c$ off-set, $T_c$ middle and $T_c$ onset of our samples read from Figure 2 are shown in Table 2.

| Compound | $T_c$ off-set (K) | $T_c$ middle (K) | $T_c$ onset (K) |
|----------|-------------------|------------------|-----------------|
| Y123 | 87 | 93.8 | 97 |
| Y5-6-11 | 96 | 102.2 | 105 |
| Y7-9-16 | 87 | 98.3 | 103 |
| Y358 | 87 | 94.3 | 97 |
| Y5-8-13 | 80 | 91.8 | 96 |
| Y7-11-18 | 96 | 102.8 | 109 |
| Y156 | 87 | 92.1 | 95 |
| Y3-8-11 | 98 | 100.4 | 102 |
| Y13-20-33 | 87 | 95.5 | 99 |

Table 2. The summation of the $T_c$ of our samples.

We find that the highest $T_c$ onset is 109 K that of Y7-11-18.and Y156 is the lowest $T_c$ onset that 95 K. Y3-8-11 is the highest $T_c$ off-set, 98 K.

We preliminary survey the crystal structure of our samples. By comparing the XRD spectra from $10^0$ to $90^0$ of our new superconductors with our Y123 spectra, they shown that the main peaks are the ones which exist in Y123 with some impurities peak. We find that there is the one of main peak at about $15^0$ that do not included in the Aliabadi,Farshchi and Akhavan [7] 's calculation of Y358. We find that our samples are shown the same crystal structure of Y123 with some impurities peaks that occur by the missing of Y-atom in some planes agreed with our assumptions.

**4.Result and Discussion**

We make the assumptions to synthesize a new superconductor in YBaCuO-family as the number of Ba-atom plus Y-atom are equal to the number of Cu-atom and the number of Y-atom can be missing to reach the higher $T_c$ but the 1st assumption must be obeyed. The new formula of YBaCuO superconductors are synthesized by using the standard solid state reaction method as Y5-6-11, Y7-9-16, Y5-8-13, Y7-11-18, Y156, Y3-8-11, Y13-20-33. The Y 7-11-18 has the highest $T_c$ onset as 109 K. Our samples are shown the same crystal structure as Y123 with some impurities peaks that occur by the missing of Y-atom in some planes agreed with our assumptions.

The effect of oxygen-doping on superconductors do not consideration in this paper that may be the one of the main parameters to increase the critical temperature. We think that the highest $T_c$ superconductor may be found by using our assumptions and the optimize doping. The more experimental detail will reveal the mechanism of occurring superconductivity in this material.

**5.Conclusions**

We find the assumptions to synthesize a new superconductor in YBaCuO-family with the higher $T_c$. The new formula of YBaCuO superconductors are synthesized by using the standard solid state reaction method as Y5-6-11, Y7-9-16, Y5-8-13, Y7-11-18, Y156, Y3-8-11, Y13-20-33. The Y 7-11-18 shown highest $T_c$ onset as 109 K Our samples are shown the same crystal structure as Y123 with some impurities peaks that occur by the missing of Y-atom in some planes agreed with our assumptions.

**Acknowledgement**
The authors would like to thank Professor Dr.Suthat Yoksan for the useful discussion and also thank the Office of Higher Education
Commission, Faculty of Science Srinakharinwirot University,and ThEP Center for the financial support.

6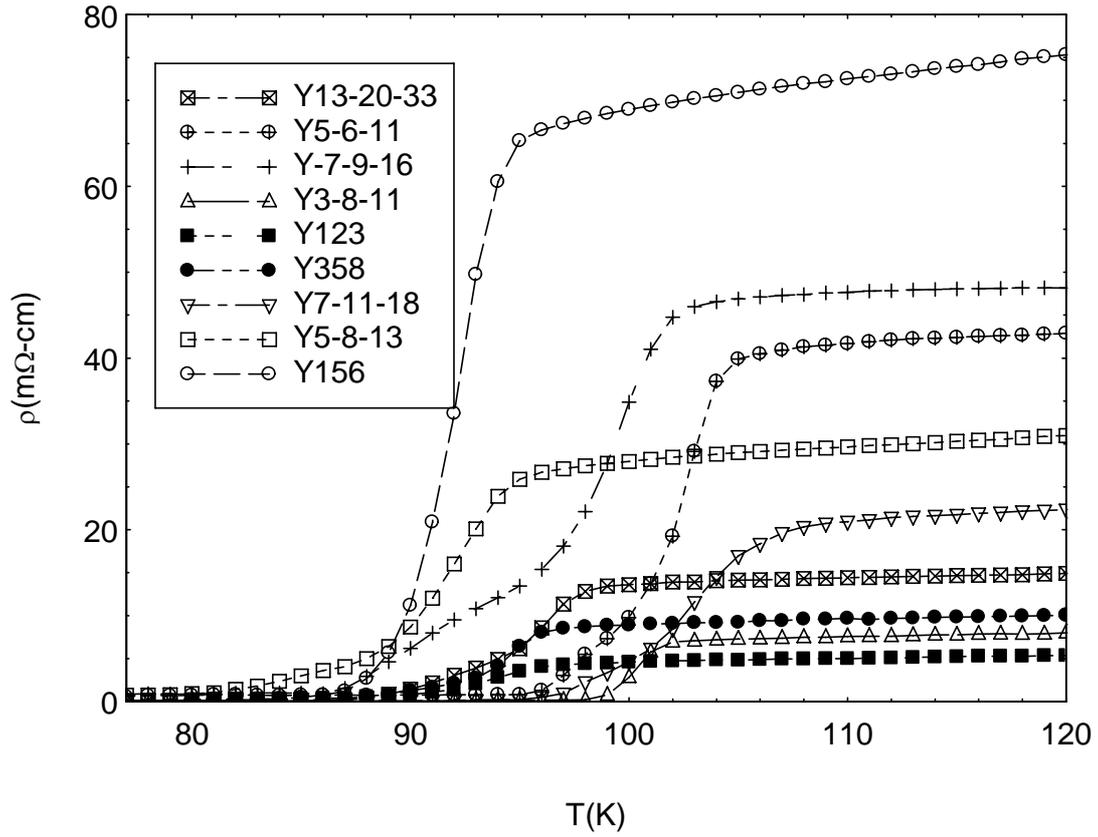

Figure 1  The resistivity versus temperature are shown.

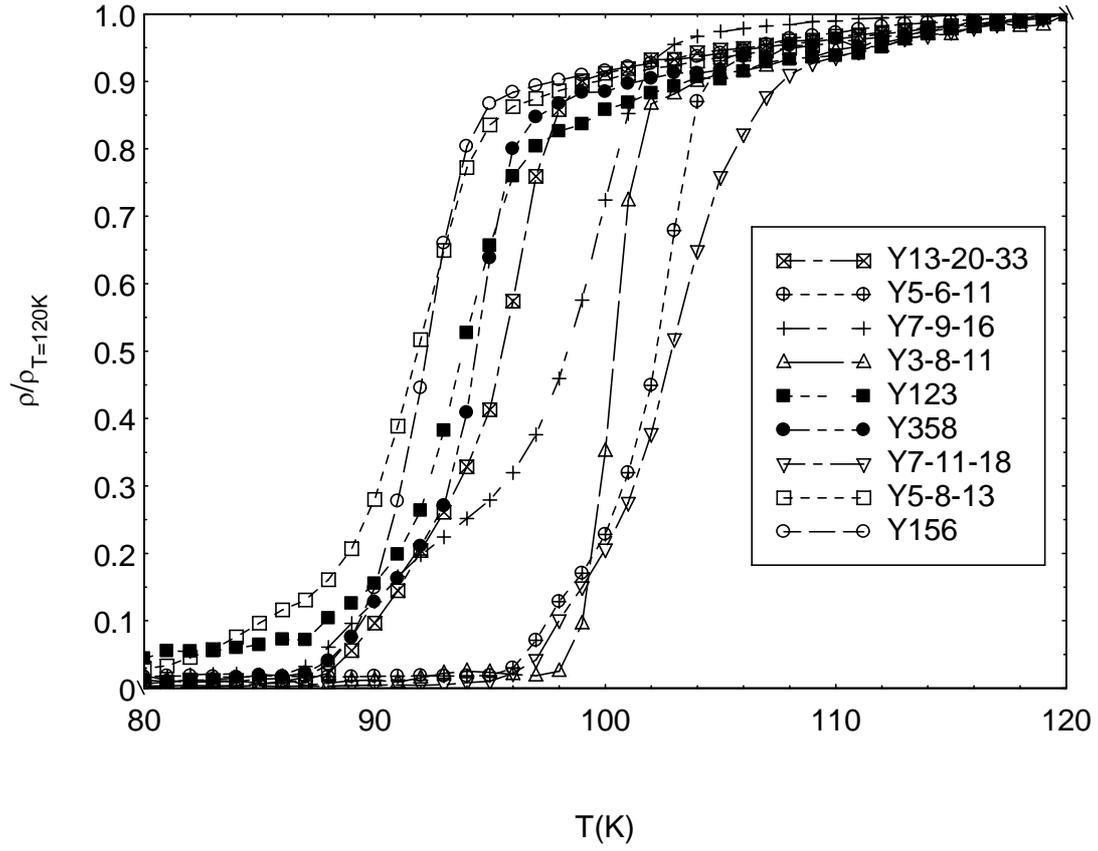

Figure 2  The normalized resistivity versus temperature are shown.